\documentclass[a4paper]{mem}
\usepackage{natbib}
\usepackage{graphicx}
\usepackage[a4paper]{hyperref}
\idline{73}{23}
\begin{document}
   \title{The Carina project:\\ 
    color magnitude diagram and radial distribution
}

   \author{M. Castellani \inst{1},
           G. Bono \inst{1},
	   R. Buonanno \inst{1},
	   F. Caputo \inst{1},
	   V. Castellani \inst{1},
	   C.E. Corsi \inst{1},
	   M. Dall'Ora \inst{1,2},
	   M. Marconi \inst{3},
	   M. Monelli \inst{1,2},
	   M. Nonino \inst{4},
	   L. Pulone \inst{1},
	   V. Ripepi \inst{3},
	   H.A.Smith \inst{5},
	   A.R. Walker \inst{6}\fnmsep 
}

   \offprints{M. Castellani}

   \institute{INAF, Osservatorio Astron. Roma,
via Frascati 33, 00040 Monteporzio Catone, Rome, Italy. \email{marco@gruppolocale.net}
\and Universit\`a di Tor Vergata, Rome, Italy
\and INAF, Osservatorio Astron. Capodimonte, Napoli, Italy
\and INAF, Osservatorio Astron. Trieste, Italy 
\and Harvard-Smithsonian Center for Astrophysics, USA
\and National Opt. Astr. Obs., Tucson, USA \\
             }

   \abstract{
   We present B, V photometric data of the Carina Dwarf Spheroidal Galaxy (dSph),
   collected with the Wide Field Imager (WFI) 
   available at the 2.2m MPI/ESO telescope.
   We briefly discuss the main features of the color magnitude diagram (CMD) 
   and in particular the mix of stellar populations present in this galaxy.
   A preliminary analysis of the spatial distribution of 
   these populations over a substantial fraction of the body of the galaxy
   is also presented.
   }
   \authorrunning{M. Castellani et al.}
   \titlerunning{The Carina project: Part Two}
   \maketitle

\section{Introduction}

%
   \begin{figure*}
    \vspace{8cm}
     \caption{(See "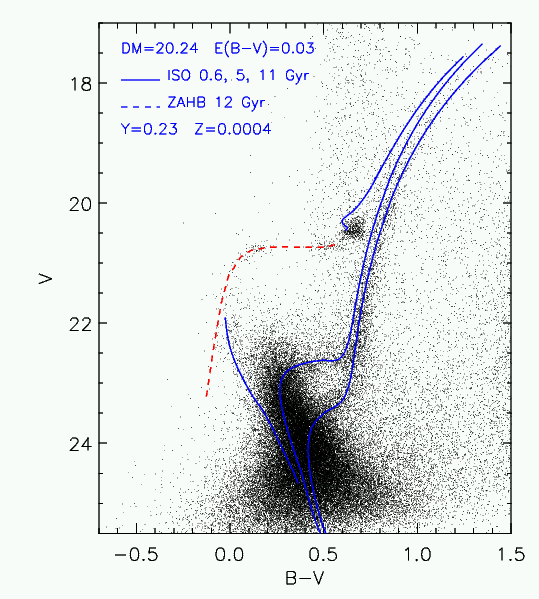") The CMD of Carina.
             Selected isochrones and a Zero Age Horizontal
             Branch (for the labeled assumption on distance modulus, reddening, and   chemical composition)
	     are superimposed on the CMD of the galaxy.
   }
              \label{FigGam}%
    \end{figure*}

DSphs are key stellar systems in
several long-standing astrophysical problems
\citep{dekel}, 
tightly connected with the
evolutionary history of Local Group (LG) galaxies
\citep{grebel}.
Empirical evidence suggest that the bulk of LG dSphs
have experienced
a much more complex star formation (SF) history than Galactic globular clusters
\citep{hodge}.
Therefore, the stellar population content of dwarf galaxies is a target
of great relevance.
However, the limited area covered by astronomical detectors hampered for a long time
an exhaustive investigation of these extended 
celestial objects and only recently the 
use of wide field detectors allows extensive 
sampling of their stellar populations. 

To take
advantage of such an opportunity we planned to use the WFI available
at 2.2m MPI/ESO telescope to investigate stellar populations in Carina.
In the dSph realm, the Carina galaxy is one of the most 
prominent example of multiple stellar populations, since its CMD 
shows a complex star-formation history. It has
been suggested by \citet{hurley} that this galaxy underwent significant bursts of star 
formation at 3, 7 and 15 Gyrs ago.

The \emph{Carina Project} was driven by these stimulating peculiarities and aims at
fully exploiting modern technological capabilities of wide field imagers
to improve current knowledge of the 
stellar content of LG dSphs. In the following we present preliminary 
results concerning the CMD and the radial distribution of the Carina stellar
populations.

More detailed information concerning \emph{variable stars}
and \emph{evolutionary properties} can be found in the papers by
Dall'Ora et al. and by Monelli et al. in these proceedings.


\section{Observations and data reduction}

Multiband B and V time series data of the 
Carina dSph were collected over three consecutive nights.
To allow the sampling of the light curve of radial variables such as RR Lyrae and
Anomalous Cepheids, we
secured 54 consecutive B and V 
exposures of $\simeq$ 500 sec. each. The observed field covered an area
of 34' x 33', pointed at the center of the galaxy. We have had an average seeing $\leq$ 1.0 arcsec both in
the B and V band. This allowed us to obtain accurate photometry (S/N$\simeq$50) down to V$\simeq$23
and B$\simeq$23.5

\emph{NOAO mscred} tasks (in the \emph{IRAF} data analysis environment) were used for bias subtraction, dark count
correction and flat fielding. With the exception of a few low-quality images, we stacked the eight individual
CCD chips, using the entire set of B and V images (we used the \emph{DAOPHOT} procedure \emph{daomatch}, 
\emph{daomaster} and \emph{montage}).
Photometry over the median image was carried out by using DAOPHOT package.
A total of 68000 stars were identified and measured with
the PSF fitting algorithm \emph{allstar}.


\section {The color magnitude diagram}

Fig. 1 shows the CMD of the galaxy, obtained following
the procedure outlined above. Main features of the diagram are: 
(i) broad main sequence, (ii) multiple, well-populated sub-giant
branches, (iii) sizeable samples of blue and red horizontal branches stars,
(iv) a well defined plume of blue, bright stars (V$\simeq$22).

To provide quantitative estimates of both mean metallicity and distance modulus of  
Carina, we superimposed selected stellar tracks and isochrones, from the 
"Pisa Evolutionary Group" (Castellani et al. 2003, in preparation) on our CMD. 
Best fit is obtained for values of metallicity Z$\simeq$0.0004 and for distance modulus
(m-M)$\simeq$20.24. As a whole, the age of the old stellar component in Carina safely
ranges from 10 to 12 Gyrs, while the bulk of "young" stars present 
turn-off ages of the order of $\simeq$5 Gyrs.

Moreover, our CMD brings forward the occurrence of younger Main Sequence (MS) stars with
an age of the order of 1 Gyrs. Such an occurrence is interesting 
because data available in the literature so far 
suggested that the most recent burst of SF
took place $\simeq$ 2-3 Gyr ago.


\section{Radial distribution}

On the basis of preliminary empirical evidence \citet{mighell}
suggested 
that the old and the "not-too-old" populations in Carina present
different radial distributions. To investigate this issue, 
we selected stars in suitable 
boxes representative of both populations (see Fig.2). 
When compared to similar analyses available in the literature, our approach
presents the substantial advantage of quite large stellar samples, 
both for "old" ($\simeq$1000)
and "young" ($\simeq$5000) populations.


\begin{figure}[b]
\includegraphics[width=6.5cm,clip]{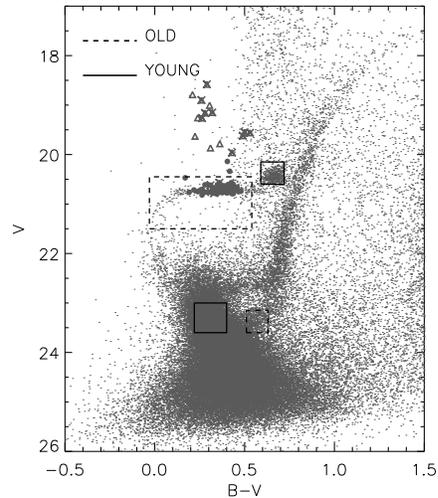}
\caption[]{Dashed and solid boxes show the CMD regions selected as 
representative of the "old" and the "young" population
respectively. Circles are RR Lyrae, triangles, AC.
Crosses mark variables with poor-phase coverage.}
\label{count}
\end{figure}


Fig. 3 (left panel) shows that there is a detectable offset, of about two arcmin, in the peack 
density between the "old" population and the Carina center. On the other end, 
we found that (Fig.3, right
panel) the isodensity contours of the "young" population appear much more regular
when compared to the "old" one, resembling quite well the density distribution 
of the entire galaxy.


   \begin{figure*}
   \centering
   \resizebox{\hsize}{!}{\includegraphics[clip=true]{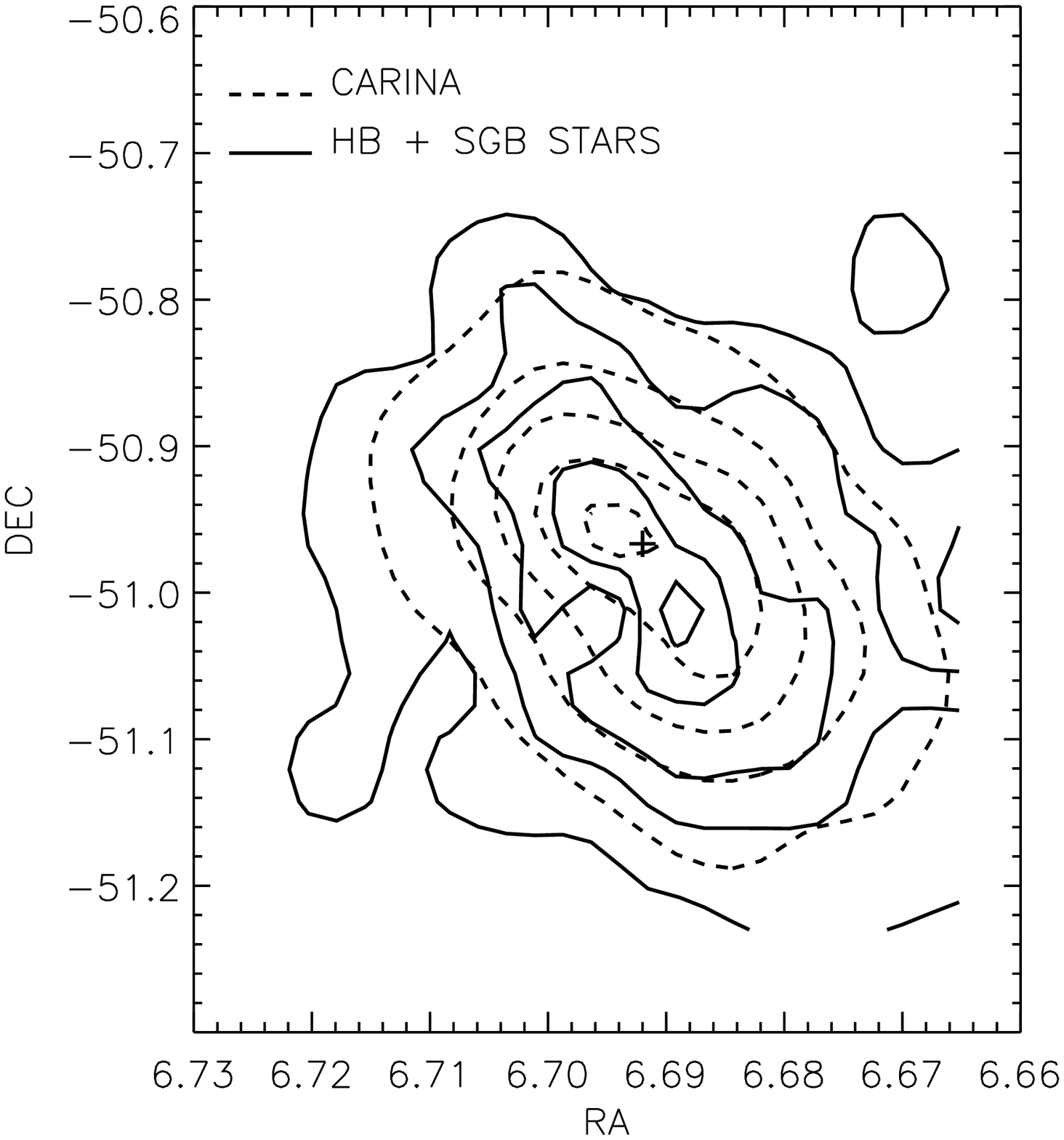}
   \includegraphics[clip=true]{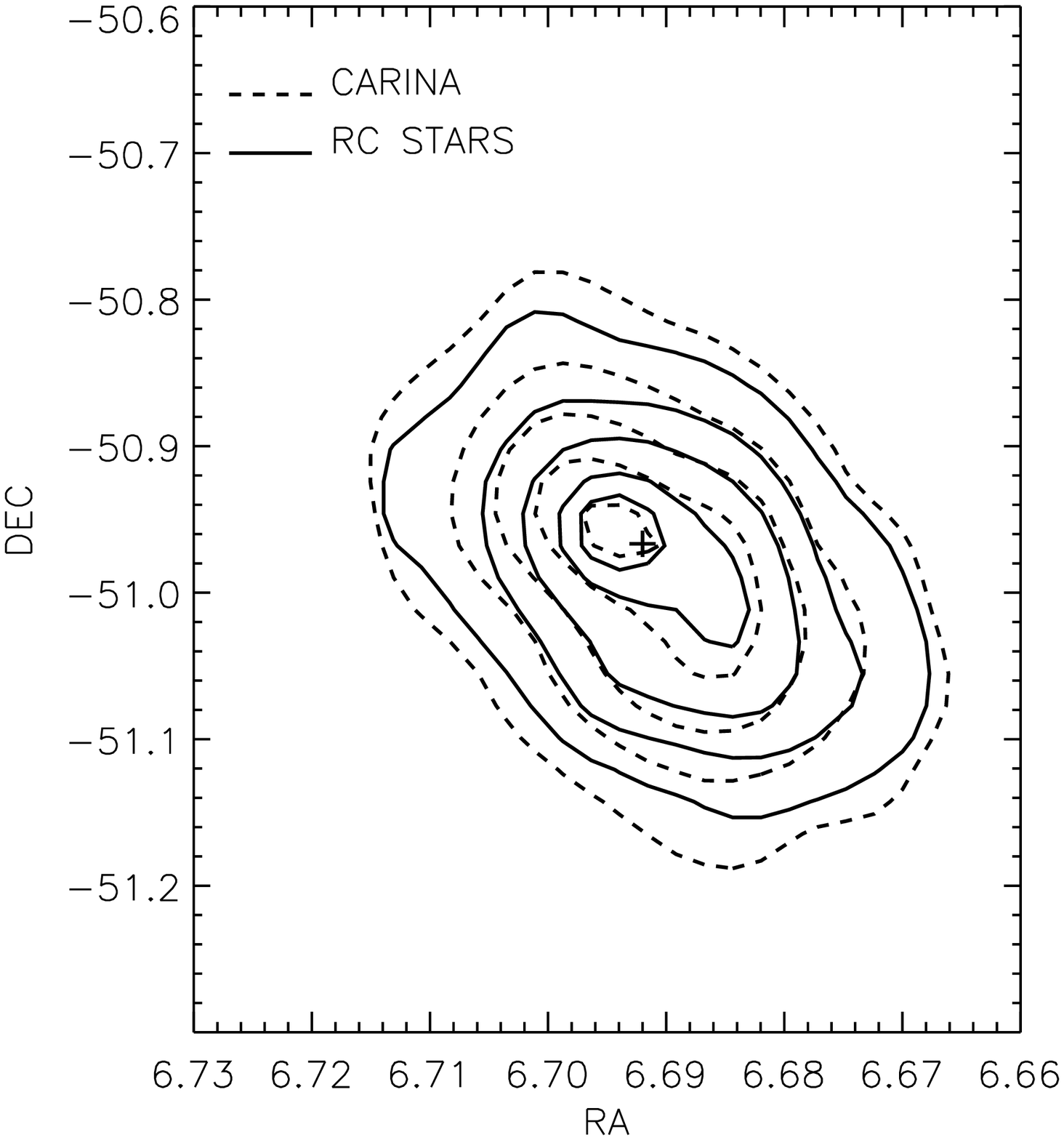}}
     \caption{Left panel: Isodensity map of the Carina stellar content (dashed)
            and of the "old" stellar population (solid)  
   	    Right panel: Isodensity map of the Carina stellar content 
	    (dashed) and of the "young" stellar population (solid). In both 
            panels, a cross
            marks the center of the galaxy. 
               }  
        \label{count}
    \end{figure*}


To estimate on a quantitative basis the difference between the two populations we performed several 
Kolmogorov-Smirnov (KS) tests. We found that the two "old" subsamples (Horizontal Branch, HB,   
and Sub Giant Branch, SGB) present the same
radial distribution, since the KS probability is equal to $\simeq$ 90\%.
The same outcome applies to the two "young" samples, namely Red Clump 
stars and not-too-old MS, and indeed
the probability given by the KS test is as hight as $\simeq$ 94\%.
On the other hand, the KS test applied to the "old" and the "young" samples, supplies a vanishing
probability that the two distribution are the same.

We also investigated the spatial distributions of the two populations along the major and minor axis. 
We found that the distribution of the "old" component is broader than the distribution of the 
"young" component. \emph{These evidence further support the suggestion of a difference in the spatial 
distribution between the "old" and the "young" stellar populations.}


\section{Conclusions}

We performed photometry of Carina DSph using the 2.2 WFI at ESO.
Our investigation shows several interesting features both for the CMD and for the
radial distribution of the stars. Moreover, current photometry suggests that last 
star formation episode took place $\simeq$ 1 Gyr
ago, at variance with previous findings in literature. 
We also found that 
\emph{stellar populations
of different age present different radial distribution:} 
the spatial distribution of the old population 
along the major and minor axes is more asymmetric when compared with the
intermediate age population.
Moreover, the peak of this population is $\simeq$ 2 arcmin off-center.


\bibliographystyle{aa}

\begin{thebibliography}{}



\bibitem [{Dekel \& Silk (1986)}]{dekel} 
Dekel, A. \& Silk, J. 1986,
ApJ, 303, 39

\bibitem [{Grebel (2001)}]{grebel} 
Grebel, E. 2001,
Astrophysics and Space Science Supplement, 277, 231

\bibitem [{Hodge (1989)}]{hodge} 
Hodge, P. 1989
Ann. Rev. Astron. Astrophys. 27, 139


\bibitem [Hurley-Keller et al. (1998)]{hurley} 
Hurley-Keller, D., Mateo, M., Nemec, J.,
1998, AJ, 115, 1840

\bibitem [Mighell (1997)]{mighell}
Mighell, K.J.
1997, AJ, 114, 1458  



\end{thebibliography}

\end{document}